# Chiral Landau levels induced by two in-plane pseudomagnetic fields in underwater acoustic metamaterials

Jiao Shen [a], Zhiyong Chang [a, c], Xinzong Wang [a], Cheng Lin [a], Tuo Liu [d], Xiaoxiao Wu [e], Yifan Zhu [a, b], Haiyan Fan [a, b, *], Hui Zhang [a, b, *]

[a] *Jiangsu Key Laboratory for Design and Manufacturing of Precision Medicine Equipment, School of Mechanical Engineering, Southeast University, Nanjing 211189, China*

[b] *Key Laboratory of Underwater Acoustic Signal Processing (Southeast University), Ministry of Education, Nanjing, 210096, China*

[c] *Advanced Ocean Institute of Southeast University Nantong, Nantong 226010, China*

[d] *State Key Laboratory of Acoustics and Marine Information, Institute of Acoustics, Chinese Academy of Sciences*

[e] *Modern Matter Laboratory and Advanced Materials Thrust, The Hong Kong University of Science and Technology (Guangzhou), Nansha, Guangzhou 511400, Guangdong, China*

**Abstract**

The chiral zeroth Landau levels (LLs) constitute topologically protected bulk states that enable robust control of acoustic wave propagation. Given the central role of underwater acoustics in marine engineering, realizing such Landau-level physics in underwater acoustic systems is highly desirable. Nevertheless, existing studies have primarily been limited to airborne acoustic systems, and the implementation of chiral zeroth LLs in underwater acoustics remains a challenge due to the unavoidable fluid-solid interactions. In this study, we realize two kinds of chiral LLs in an open underwater spoof surface acoustic wave (SSAW) platform by introducing two perpendicular in-plane artificial pseudomagnetic fields (PMFs), oriented along the $x$ and $y$ directions, respectively, and reveal that scalar acoustic fields in water and vectorial elastic vibrations in solids can be jointly manipulated within a unified framework. Specifically, by strategically opening bandgaps at the Dirac points, position-dependent effective mass terms are introduced into the Dirac Hamiltonians, thereby synthesizing two in-plane PMFs. This results in the emergence of chiral LLs, which is confirmed both numerically and experimentally. The unidirectional propagation of the chiral LLs and their robustness against defects are also demonstrated. In addition, we achieve flexible manipulation of underwater ultrasonic energy carried by SSAWs, including beam splitting and arbitrary wave steering. Dual-band chiral LLs are also observed in small-scale underwater topological metamaterials. Our work provides a new route toward SSAW-based underwater ultrasonic control, opening opportunities for multiband underwater acoustic signal processing and detection, as well as underwater acoustic energy harvesting.

## 1 Introduction

Oceans occupy more than 70% of the Earth's surface and contain abundant natural resources. As electromagnetic waves attenuate rapidly in water, acoustic waves are an important tool in fields such as underwater wireless communication and target detection, which necessitates the flexible manipulation of waterborne acoustic waves [1,2]. In recent years, underwater acoustic metamaterials have attracted considerable attention as they can enable various acoustic wave manipulations that are difficult to achieve using natural materials, including invisibility cloaking [3–5], beam formation [6–9], phase engineering [10–15], topological acoustics [16–24], and sound absorption [25–28]. Among them, studies in underwater topological acoustics primarily focus on the realization of the quantum spin Hall-like effect [16,17], quantum valley Hall effect [18,19], and higher-order topological insulator [20,21]. However, due to the limited acoustic impedance contrast between water and solids, the interactions of sound in water and solids become considerable. As a result, research progress on underwater acoustic metamaterials lags far behind that on airborne acoustic metamaterials. One typical example is the Landau levels (LLs) induced by the underwater pseudomagnetic fields (PMFs).

* Corresponding author.
*E-mail address:* jiaoshen@seu.edu.cn (J. Shen), zhiyongchang@seu.edu.cn (Z. Chang), 18816203250@163.com (X. Wang), Chenglin@seu.edu.cn (C. Lin), liutuo@mail.ioa.ac.cn (T. Liu), xiaoxiaowu@hkust-gz.edu.cn (X. Wu), yifanzhu@seu.edu.cn (Y. Zhu), haiyan.fan@seu.edu.cn (H. Fan), seuzhanghui@seu.edu.cn (H. Zhang)

Unlike edge states in quantum (spin/valley) Hall effect and corner states in higher-order topological insulators, topological LLs represent a class of topologically protected bulk states. LLs, originally predicted and subsequently observed in condensed matter systems, arise from the quantization of electronic cyclotron motion under external magnetic fields [29–32]. The realization of LLs in acoustic systems was long impeded by the spinless and magnetically insensitive nature of acoustic waves, and has become feasible through the introduction of strain-induced out-of-plane PMFs [33–39]. Displacement of the Dirac points in two-dimensional (2D) momentum space produces such out-of-plane PMFs. The resulting Landau bands exhibit flat dispersion with zero group velocity [40–44]. Apart from flat Landau bands, chiral LLs have been discovered in three-dimensional (3D) acoustic and photonic Weyl semimetals. These chiral LLs possess non-zero group velocity and form topologically protected one-way bulk states, thereby laying the foundation for chiral transmission [45–49]. Relative to 3D Weyl systems with complicated configurations, the 2D counterparts with simpler geometrical structures offer broader application prospects, directing increasing studies toward 2D systems. To date, chiral LLs realized by synthesizing in-plane PMFs have been experimentally demonstrated in photonic [50–52], airborne acoustic [53], solid acoustic [54–58], and electromagnetic metasurfaces [59–61] systems, but remain unexplored in underwater acoustic metamaterials involving fluid-solid interactions.

In this work, chiral LLs of underwater spoof surface acoustic waves (SSAWs) are theoretically and experimentally demonstrated in an open fluid-solid coupled platform. Compared with enclosed airborne acoustic waveguides, this open platform offers greater potential for practical applications. In particular, in-plane PMFs along the $x$ and $y$ directions, denoted by $B_x$ and $B_y$, respectively, are designed to realize chiral LLs. $B_\mathrm{x}$ is synthesized by strategically opening local bandgaps at Dirac points to induce a $y$-dependent effective mass, while $B_\mathrm{y}$ is realized analogously by introducing an $x$-dependent effective mass. We experimentally measure the band structures of the zeroth LLs, demonstrating that zeroth LLs with negative group velocity belong to the $K'$ valley, which is consistent with the theoretical results. The emergence of these zeroth LLs in fluid-solid coupled systems further substantiates the possibility of simultaneously controlling scalar acoustic fields and vectorial elastic vibrations within a unified framework. Moreover, the unidirectional transmission of underwater ultrasound mediated by these chiral LLs is observed, together with their robustness against defects. Building upon this, flexible control of underwater ultrasound is achieved, enabling beam splitting and wave steering along arbitrary trajectories. Dual-band chiral LLs in a small-scale underwater acoustic metamaterial are also investigated. These findings highlight the advantages of chiral LLs in an open underwater SSAW platform for robust ultrasonic manipulation.

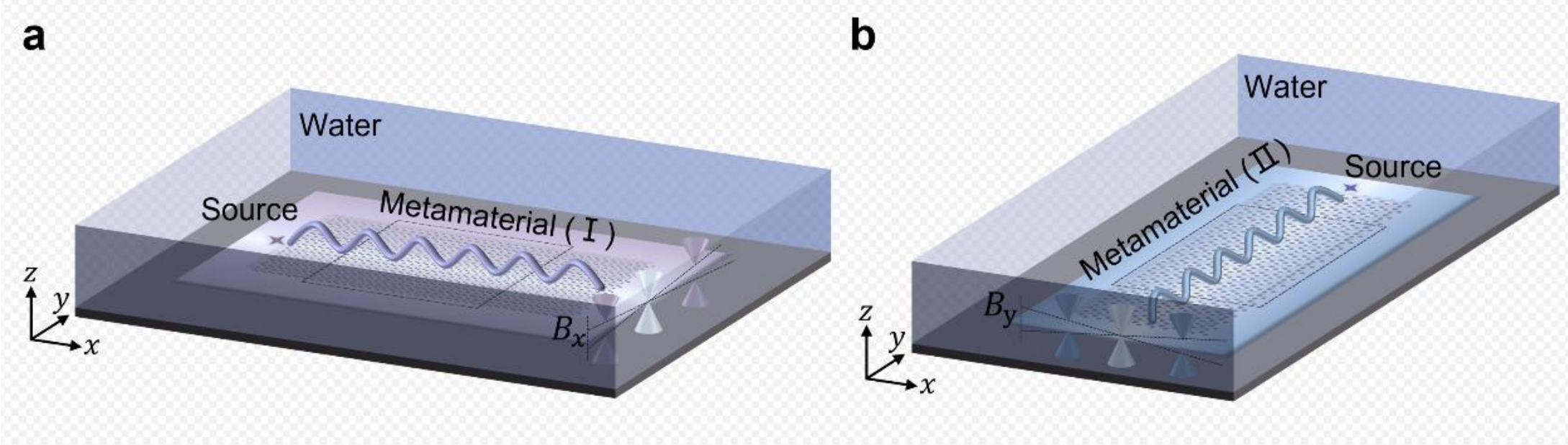

**Fig. 1.** Conceptual illustrations of two valley-dependent chiral LLs induced by in-plane PMFs along the $x$ and $y$ directions, namely (a) $\boldsymbol{B}_x$ and (b) $\boldsymbol{B}_y$, respectively, in underwater acoustic metamaterials.

## 2 Theory

We begin by describing the principles underlying the synthesis of the in-plane PMFs and the process of realizing chiral Landau quantization. Figure 1 presents two distinct microperforated phononic-plate structures designed to generate two in-plane PMFs, each with a plate thickness of $l = 1.5$ mm. The plate is made of aluminum, with an elastic modulus of 70 GPa, a density of 2700 kg/m$^3$, and a Poisson's ratio of 0.33. The structure is based on an acoustic honeycomb lattice composed of circular holes immersed in water, supporting underwater SSAWs localized near the fluid-solid interface. The corresponding unit cell, shown in Fig. 2(a), consists of circular holes at sublattices A and B, with hole diameters of $d_A = d_B = 1$ mm and a lattice constant of $a = 2.598$ mm. The structure possesses $C_{6v}$ symmetry, which supports Dirac cones at the $K/K'$ points of the Brillouin zone. By tuning the diameter difference ($\Delta d = d_A - d_B$) between the circular holes at sublattices A and B, inversion symmetry is broken and a bandgap is opened at the Dirac point, as displayed in Fig. 2(b).

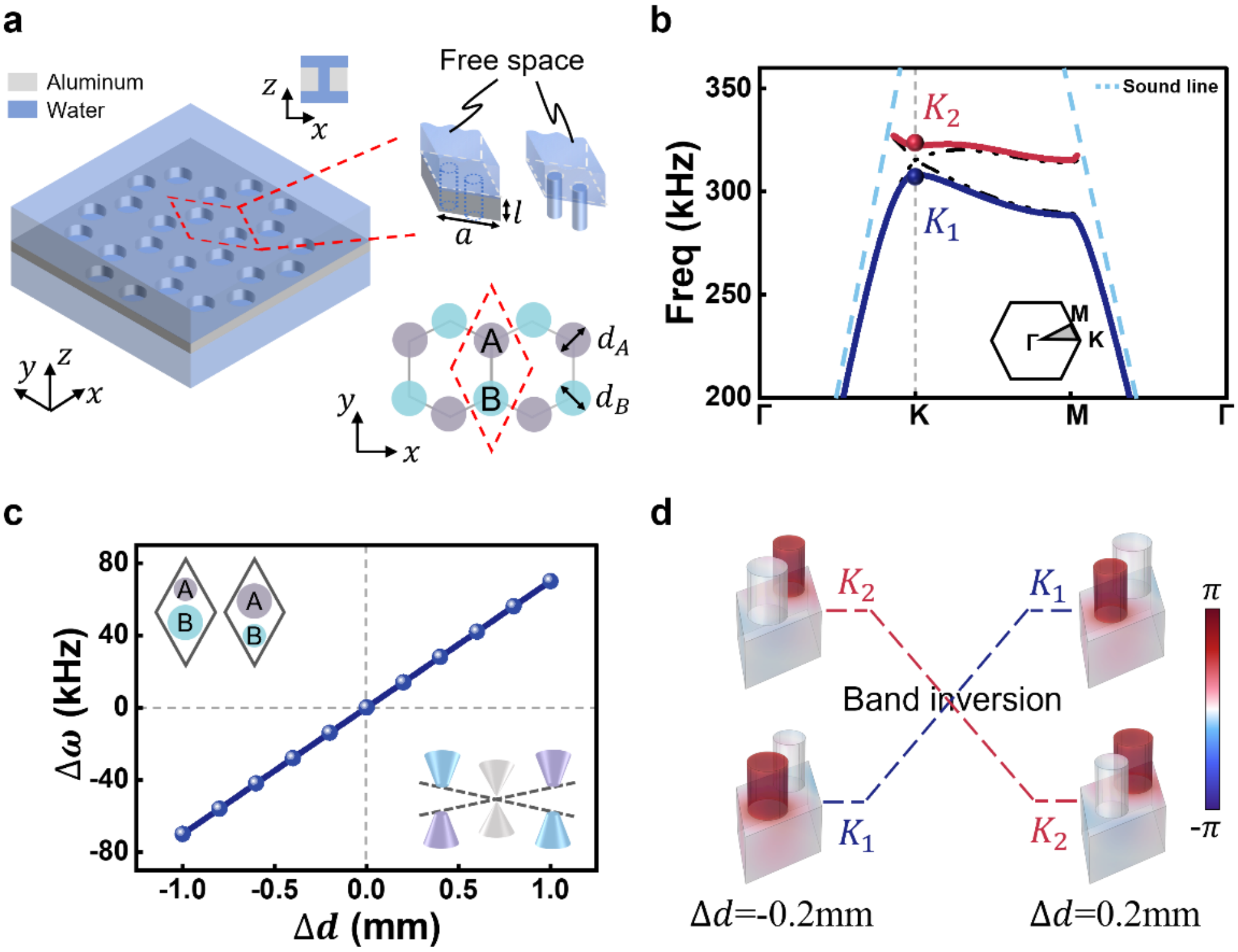

**Fig. 2.** Unit-cell design and band inversion in the underwater SSAW metamaterial. a) Perspective view of the microperforated aluminum plate immersed in water and schematic of the honeycomb-lattice unit cell, in which sublattices A and B are formed by circular through holes with diameters $d_A$ and $d_B$, respectively. The wavy surfaces represent the free-space water domains extending along the $z$ direction. b) Calculated band structures along the high-symmetry path (Γ-K-M-Γ). The inversion-symmetric unit cell with ($d_A = d_B = 1$ mm) exhibits a Dirac cone at the $K$ point, as indicated by the black dashed curves, whereas inversion symmetry breaking with ($d_A = 0.9, d_B = 1.1$ mm) opens a bandgap, as shown by the solid curves. The light-blue dashed curves denote the sound line. c) Variation of the unit-cell bandgap ($\Delta\omega$) with the diameter difference ($\Delta d = d_A - d_B$). The insets illustrate two inversion-related unit cells and the corresponding bandgap opening and reopening process. d) Phase distributions of the $K_1$ and $K_2$ eigenmodes for the unit cell ($d_A = 0.9, d_B = 1.1$ mm) and its inversion-related counterpart ($d_A = 1.1, d_B = 0.9$ mm), showing the band inversion process.

Based on the $k \cdot p$ method, the wave equation in the vicinity of the Dirac cone at the $K/K'$ points can be written as $H_D\psi = \omega_D\psi$, with $\psi$ and $\omega_D$ indicating the wave function and the Dirac frequency, respectively. In 2D systems protected by parity inversion symmetry, the Hamiltonian around the $K/K'$ points can be represented as the standard massless Dirac Hamiltonian:

$$H_D = v_D\left(\hat{k}_x\sigma_x + \hat{k}_y\sigma_y\right), \tag{1}$$

where $v_D$ is the group velocity, $\hat{k}_i(i = x, y)$ are the wave-vector operators in the $x$ and $y$ directions, and $\sigma_i(i = x, y)$ are the corresponding Pauli matrices. In electronic systems, an external magnetic field quantizes the cyclotron motion of charged particles, giving rise to LLs. However, in acoustic systems, phonons are neutral and therefore insensitive to magnetic fields, requiring the introduction of artificial PMFs. In this work, to generate artificial in-plane PMFs and achieve quantized chiral LLs, the diameter difference between the circular holes at sublattices A and B is modulated according to the spatial coordinate of each unit cell. As shown in Figs. 2(c) and 2(d), this modulation produces bandgaps of different sizes, with the local bandgap ($\Delta\omega$) varying approximately linearly with the diameter difference ($\Delta d$). Reversing the sign of $\Delta d$ interchanges the modal characteristics of $K_1$ and $K_2$, thereby inducing band inversion. Essentially, this modulation introduces unequal effective mass terms into the Dirac Hamiltonian, giving rise to in-plane PMFs.

Starting from the 2D acoustic Dirac Hamiltonian with an effective mass term, a position-dependent effective mass term $b(\boldsymbol{r})$ is introduced by locally breaking the inversion symmetry of the unit cell near the $K/K'$ points. The corresponding effective Hamiltonian can be expressed as:

$$H_D = v\left(\hat{k}_x\sigma_x + \hat{k}_y\sigma_y\right) + b(\boldsymbol{r})\sigma_z, \tag{2}$$

where $b(\boldsymbol{r})$ is described as:

$$b(\boldsymbol{r}) = \frac{\omega_{q+}(\boldsymbol{r}) - \omega_{p-}(\boldsymbol{r})}{2} = \frac{\Delta\omega(\boldsymbol{r})}{2}. \tag{3}$$

Here, $\Delta\omega$ is the band gap, fluctuating with the hole diameters ($d_A$ and $d_B$) of the unit cell, and $\boldsymbol{r} = (x, y)$ is the in-plane position vector specifying the spatial location of each unit cell in the metamaterial.

According to the substitution rule for a particle with charge of $q$ in electromagnetic fields with vector potential of $\boldsymbol{A}$, i.e., $\boldsymbol{k} \rightarrow \boldsymbol{k} + q\boldsymbol{A}$, the additional mass term $b(\boldsymbol{r})$ in the Hamiltonian is mathematically equivalent to a synthetic vector gauge potential along the out-of-plane direction, i.e., $\boldsymbol{A} = \left(0,0, A_\xi\right) = \left(0,0, b(\boldsymbol{r})\right)$. Since this out-of-plane vector potential $A_\xi$ is spatially dependent, and given the gauge field relation $\boldsymbol{B} = \nabla \times \boldsymbol{A}$, the system naturally induces a PMF $\boldsymbol{B}$ that lies within the 2D plane:

$$\boldsymbol{B} = \frac{\partial b(r)}{\partial_y}\hat{e}_x + \frac{\partial b(r)}{\partial_x}\hat{e}_y. \tag{4}$$

Therefore, the orientation of the in-plane PMF can be controlled by tailoring the spatial distribution of the macroscopic structural gradient.

### *2.1 Chiral Landau levels induced by x-directed pseudomagnetic fields*

When the gradient of the acoustic unit cells is applied exclusively along the $y$-axis, the mass term scales linearly as $b(\boldsymbol{r}) = b(y) = my$, which yields a uniform synthetic magnetic field directed along the $x$-axis ($B_x = m\hat{e}_x$). Eq. (4) can then be expressed by:

$$H_D = \begin{bmatrix} my & v(\hat{k}_x - i\hat{k}_y) \\ \hat{k}_x + i\hat{k}_y & -my \end{bmatrix}. \tag{5}$$

Left-multiplying both sides by $H_D$, the squared Hamiltonian expands as:

$$H_D^2 = \begin{bmatrix} m^2y^2 + v^2(\hat{k}_x^2 + \hat{k}_y^2) & ivm \\ -ivm & m^2y^2 + v^2(\hat{k}_x^2 + \hat{k}_y^2) \end{bmatrix}. \tag{6}$$

By utilizing the canonical commutation relations $[y, \hat{k}_y] = i$ and $[x, \hat{k}_x] = i$, the underlying eigenvalue problem is mapped onto an uncoupled effective wave equation:

$$[m^2y^2 + v^2(\hat{k}_x^2 + \hat{k}_y^2) + mv\sigma_x]\psi = \omega^2\psi. \tag{7}$$

To obtain an algebraic solution to Eq. (7), spatial annihilation and creation operators associated with the harmonic confinement along the $y$ direction are introduced:

$$a = \frac{1}{\sqrt{2v|m|}}(v\hat{k}_y + imy), a^\dagger = \frac{1}{\sqrt{2v|m|}}(v\hat{k}_y - imy). \tag{8}$$

These operators satisfy the standard bosonic commutation relation: $[a, a^\dagger] = 1$. The value of $N$ can be determined by:

$$\frac{1}{N^2}([v\hat{k}_y, i|m|y] + [-i|m|y, v\hat{k}_y]) = \frac{2v|m|}{N^2} = 1. \tag{9}$$

Thus, $N = \sqrt{2v|m|}$, and the corresponding number operator $\hat{n} = a^\dagger a$ can be expanded as:

$$\hat{n} = \frac{1}{N}(v^2\hat{k}_y^2 + v|m| + m^2y^2). \tag{10}$$

Substituting Eq. (10) back into Eq. (7) yields:

$$[N^2\hat{n} - v|m| + v^2\hat{k}_x^2 - \omega^2]\psi = -mv\sigma_x\psi. \tag{11}$$

To remove the $\sigma_x$ term, we square both sides of Eq. (11) and subsequently take the square root. Considering that the system retains translational symmetry along the $x$-axis (making $k_x$ a good quantum number), the simplified characteristic equation becomes:

$$[2|m|v\hat{n} - |m|v + v^2\hat{k}_x^2 - \omega^2]\psi = \pm mv\psi. \tag{12}$$

Squaring both sides of Eq. (12) and extracting the eigenvalues yields the quantized energy spectrum for the chiral LLs:

$$\omega_n = \begin{cases} \chi sgn(m) v k_x, n = 0 \\ \pm\sqrt{v^2 k_x^2 + 2n|m|v}, n \geq 1 \end{cases}, \tag{13}$$

where $\chi = \pm 1$ represents the valley index for the $K$ and $K'$ valleys, respectively. Eq. (13) indicates that the zeroth LL ($n = 0$) exhibits the strictly linear dispersion relation along $k_x$, establishing the one-way propagating bulk state.

*2.2 Chiral Landau levels induced by y-directed pseudomagnetic fields*

Alternatively, when the structural gradient is rotated by 90° and oriented along the $x$-axis, the position-dependent mass term becomes $b(\boldsymbol{r}) = b(x) = mx$. According to the general gauge relation in Eq. (4), this configuration directly translates into a uniform synthetic magnetic field pointing along the $y$-axis ($B_y = -m\hat{e}_y$).

Due to the algebraic symmetry of the Pauli matrices, the theoretical derivation follows that of Section 2.1. After the corresponding coordinate transformation, the squared Hamiltonian in Eq. (7) can be rewritten as:

$$[m^2 x^2 + v^2(\hat{k}_x^2 + \hat{k}_y^2) - mv\sigma_y]\psi = \omega^2 \psi, \tag{14}$$

where the cross-term flips from $mv\sigma_x$ to $-mv\sigma_y$, and the harmonic confinement is reoriented along the $x$ direction. Accordingly, the ladder operators are redefined as:

$$a = \frac{1}{\sqrt{2v|m|}}(v\hat{k}_x + imx), a^\dagger = \frac{1}{\sqrt{2v|m|}}(v\hat{k}_x - imx). \tag{15}$$

Since the translational symmetry is now preserved along the $y$-axis, $k_y$ becomes a good quantum number. By diagonalizing the shifted system, the $k_y$-dispersive chiral LLs are analytically obtained as:

$$\omega_n = \begin{cases} \chi sgn(m) v k_y, n = 0 \\ \pm\sqrt{v^2 k_y^2 + 2n|m|v}, n \geq 1 \end{cases}. \tag{16}$$

In summary, the unidirectional propagation path of the topological bulk state can be flexibly routed along either the $x$ or $y$ direction by tailoring the orientation of the macroscopic lattice gradient, providing a robust mechanism for underwater sound transmission.

## 3 Results

Based on the above theory and the distinctive properties of the unit cell, two underwater acoustic metamaterial configurations are designed to generate in-plane PMFs along the $x$ and $y$ directions, respectively. These configurations are referred to as Type I ($B_x$) and Type II ($B_y$), as depicted in Fig. 1.

*3.1 Type I ($B_x$) underwater acoustic metamaterial*

The detailed structure of the Type I underwater SSAW metamaterial is illustrated in Fig. 3(a). Quantized LLs require a constant $x$-directed in-plane PMF ($B_x$) as discussed in Section 2.1. The bandgap ($\Delta\omega$) of the unit cell is therefore engineered to vary approximately linearly with the $y$ coordinate. The bandgap at the Dirac point closes and reopens as the diameter difference passes through zero ($\Delta d = 0$). This spatial modulation introduces a position-dependent effective mass term ($b(\boldsymbol{r}) = b(y) = m_1 y$), corresponding to an $x$-directed PMF ($B_\mathrm{x} = \partial b(\boldsymbol{r})/\partial_y = m_1$). A total of 21 unit cells are constructed by varying $\Delta d$, with $\Delta\omega$ increasing linearly from zero to $\Delta\omega_{max}$ along the $+y$ direction in steps of ($\Delta\omega_{max}/10$). Inversion-symmetric counterparts are arranged along the $-y$ direction, preserving translational symmetry along the $x$ direction (see Supplementary Note 1 for specific geometric dimensions).

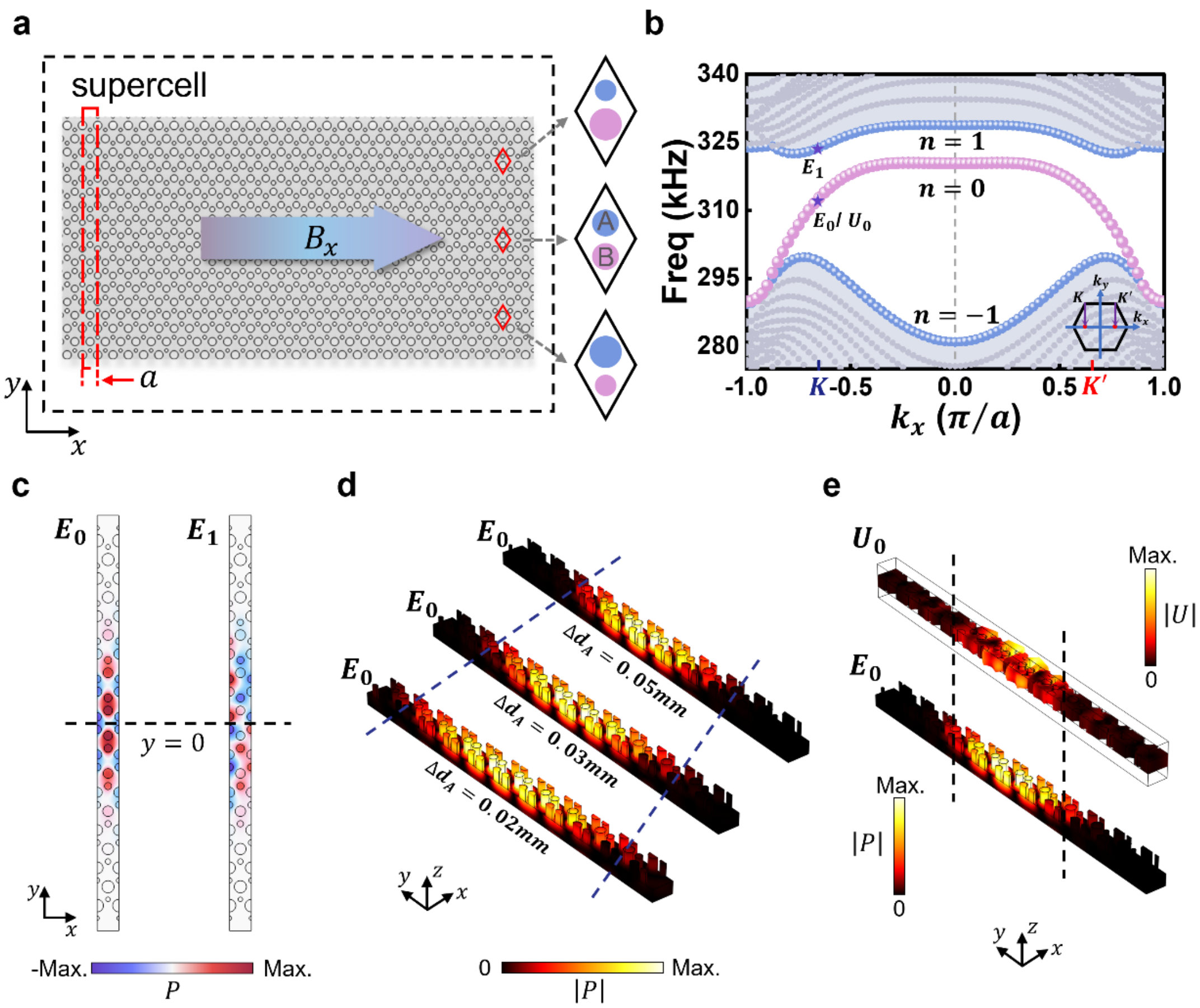

**Fig. 3.** Emergence of chiral LLs in the Type I underwater SSAW metamaterial. a) The detailed structure of the Type I underwater acoustic metamaterial. The red dashed box indicates the supercell used for the projected-band calculations, and three representative unit cells are shown on the right. b) Projected band structure of the supercell indicated in (a), with periodic boundary conditions imposed along the $x$ direction. The eigenstates $E_0/U_0$ and $E_1$ are marked in the band structure. c) Acoustic pressure fields of the eigenmodes denoted by $E_0$ and $E_1$ in (b). d) Field confinement of the zeroth LL tuned by the PMF ($B_\mathrm{x}$), whose strength is determined by the gradient ($\Delta d_A$). e) Comparison of the displacement field ($U_0$) and the corresponding acoustic pressure field ($E_0$) of the zeroth LL.

Figure 3(b) presents the projected band structure of the supercell marked by the red dashed box in Fig. 3(a), with periodic boundary conditions applied along the $x$ direction. Quantized LLs emerge in the projected spectrum, with the zeroth LL exhibiting a relatively broad bandwidth. The $x$-directed in-plane PMF projects the $K$ and $K'$ valleys onto different positions along $k_x$, resulting in the two chiral LLs not intersecting in the projected band structure. This projection process is schematically depicted in the inset of Fig. 3(b). The eigenmodes of the zeroth and first LLs, denoted by $E_0$ and $E_1$, are displayed in Fig. 3(c). The acoustic pressure distributions of $E_0$ and $E_1$ are symmetric and antisymmetric with respect to the horizontal central axis ($y = 0$) of the supercell, respectively.

The effect of the PMF strength is examined through the field distributions in Fig. 3(d), corresponding to $\Delta d_A = 0.02, 0.03,$ and $0.05$ mm. As $\Delta d_A$ increases, the high-intensity region of the zeroth LL becomes progressively narrower, indicating stronger field confinement. Here, $\Delta d_A$ denotes the difference between the diameters of the A-sublattice holes in two adjacent unit cells along the $y$ direction. According to the wave function of the zeroth LL, its localization is controlled by the PMF, whose strength is governed by the spatial gradient of the position-dependent effective mass term. Since the effective-mass gradient is closely related to $\Delta d_A$, the three values of $\Delta d_A$ correspond to calculated PMF magnitudes of $|B_x| = 14/15a_0$, $7/5a_0$, and $7/3a_0$, respectively. A larger $\Delta d_A$ therefore produces a stronger $|B_x|$, confining the energy distribution to a smaller region. This trend is consistent with the theoretical discussion in Section 2.1. As shown in Fig. 3(e), the acoustic pressure field ($E_0$) in water and the elastic displacement field ($U_0$) in the phononic crystal plate are localized within approximately the same spatial region. Both fields correspond to the same zeroth Landau eigenstate at the $K$ point, with ($E_0$) and ($U_0$) representing its acoustic pressure and elastic displacement components, respectively. This spatial co-localization reveals the hybrid fluid-solid nature of the underwater SSAW mode and indicates that acoustic pressure fields in water and elastic vibrations in solids can be controlled within a unified physical framework.

### *3.2 Type II ($B_y$) underwater acoustic metamaterial*

Different from the Type I structure, the local bandgap of each unit cell in the Type II underwater acoustic metamaterial is designed to vary approximately linearly with the $x$ coordinate (see Supplementary Note 2 for specific geometric dimensions), as shown in Fig. 4(a), yielding a constant $y$-directed in-plane PMF ($B_y$). The corresponding effective mass term can be written as $b(\boldsymbol{r}) = b(x) = m_2 x$, which gives a $y$-directed PMF ($B_y = \partial b(\boldsymbol{r})/\partial_x = m_2$). Translational symmetry is preserved along the $y$ direction, with a single period of $\sqrt{3}a$. Figure 4(b) shows the projected band structure of the supercell indicated by the red dashed box in Fig. 4(a). The zeroth LL with a relatively broad bandwidth is also observed. When the in-plane PMF is oriented along the $y$ direction, the $K$ and $K'$ valleys are projected onto the same position along $k_y$, causing the two chiral LLs to intersect in the projected band structure, as shown in the inset of Fig. 4(a). The eigenmodes of the zeroth and first LLs, denoted by $E_0$ and $E_1$, are illustrated in Fig. 4(c). The acoustic pressure distributions of $E_0$ and $E_1$ are centrosymmetric and anti-centrosymmetric with respect to the central point (0,0) of the supercell, respectively. The confinement of the zeroth LL can likewise be tuned by the PMF ($B_y$), as shown in Fig. 4(d). The gradients ($\Delta d_A = 0.02, 0.03,$ and $0.04$ mm) correspond to calculated PMF magnitudes of $|B_y| = 14/5a$, $21/5a$, and $28/3a$, respectively, with increasing $|B_y|$ leading to progressively stronger field confinement. Figure 4(e) further shows the co-localization of the acoustic pressure field in water and the elastic displacement field in the phononic crystal plate.

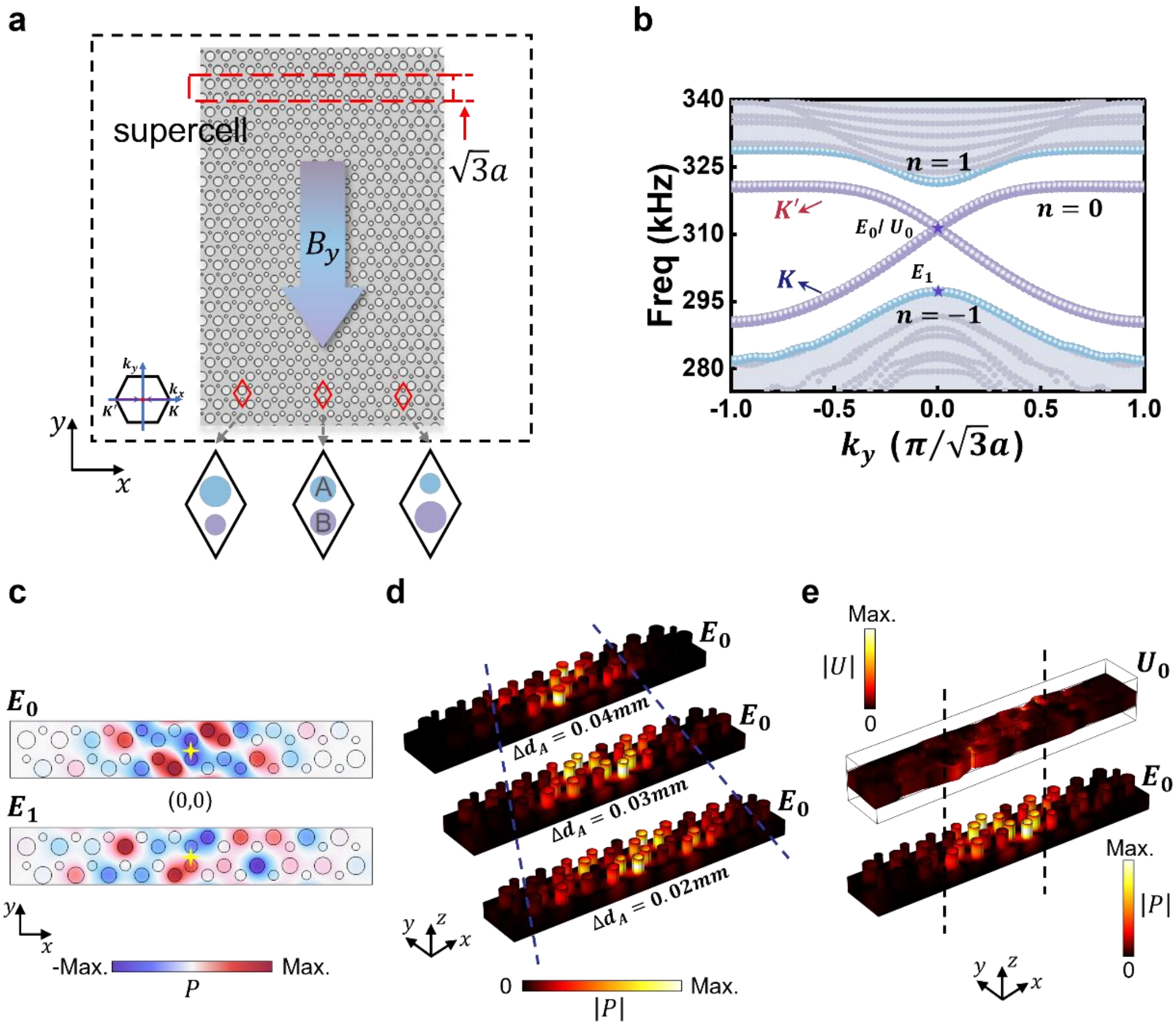

Fig. 4. Emergence of chiral LLs in the Type II underwater SSAW metamaterial. a) The detailed structure of the Type II underwater acoustic metamaterial. The red dashed box indicates the supercell used for the projected-band calculations, and three representative unit cells are shown below. b) Projected band structure of the supercell indicated in (a), with periodic boundary conditions imposed along the $y$ direction. The eigenstates $E_0/U_0$ and $E_1$ are marked in the band structure. c) Acoustic pressure fields of the eigenmodes denoted by $E_0$ and $E_1$ in (b). d) Field confinement of the zeroth LL tuned by the PMF ($B_y$), whose strength is determined by the gradient ($\Delta d_A$). e) Comparison of the displacement field ($U_0$) and the corresponding acoustic pressure field ($E_0$) of the zeroth LL.

The zeroth LLs around the $K$ and $K'$ points are valley dependent, with the $K$ valley corresponding to positive group velocity and the $K'$ valley corresponding to negative group velocity. Selective excitation of these valley states enables the propagation direction of the zeroth LL to be controlled. Both the Type I and Type II metamaterials support unidirectional propagation excited either by a chiral source placed near the center of the structure or by a point source located at the midpoint of the corresponding boundary. As representative examples, Figs. 5(a) and 5(b) show the leftward and rightward one-way propagation induced by a chiral source in the Type I metamaterial, whereas Figs. 5(c) and 5(d) demonstrate upward and downward one-way propagation excited by point sources placed at the bottom and top boundaries of the Type II metamaterial, respectively. Since unidirectionality is less visually evident when the source is positioned at a structural boundary, the acoustic pressure fields in Figs. 5(c) and 5(d) are further analyzed using 2D fast Fourier transform (FFT). The Fourier spectra in Figs. 5(e) and 5(f) show dominant excitations around the $K$ and $K'$ valleys for the upward and downward propagating waves, respectively, confirming their valley-dependent positive and negative group velocities in agreement with the projected band structure in Fig. 4(b). These results demonstrate that the propagation direction of underwater ultrasound can be selectively controlled through valley-selective excitation of the chiral zeroth LLs.

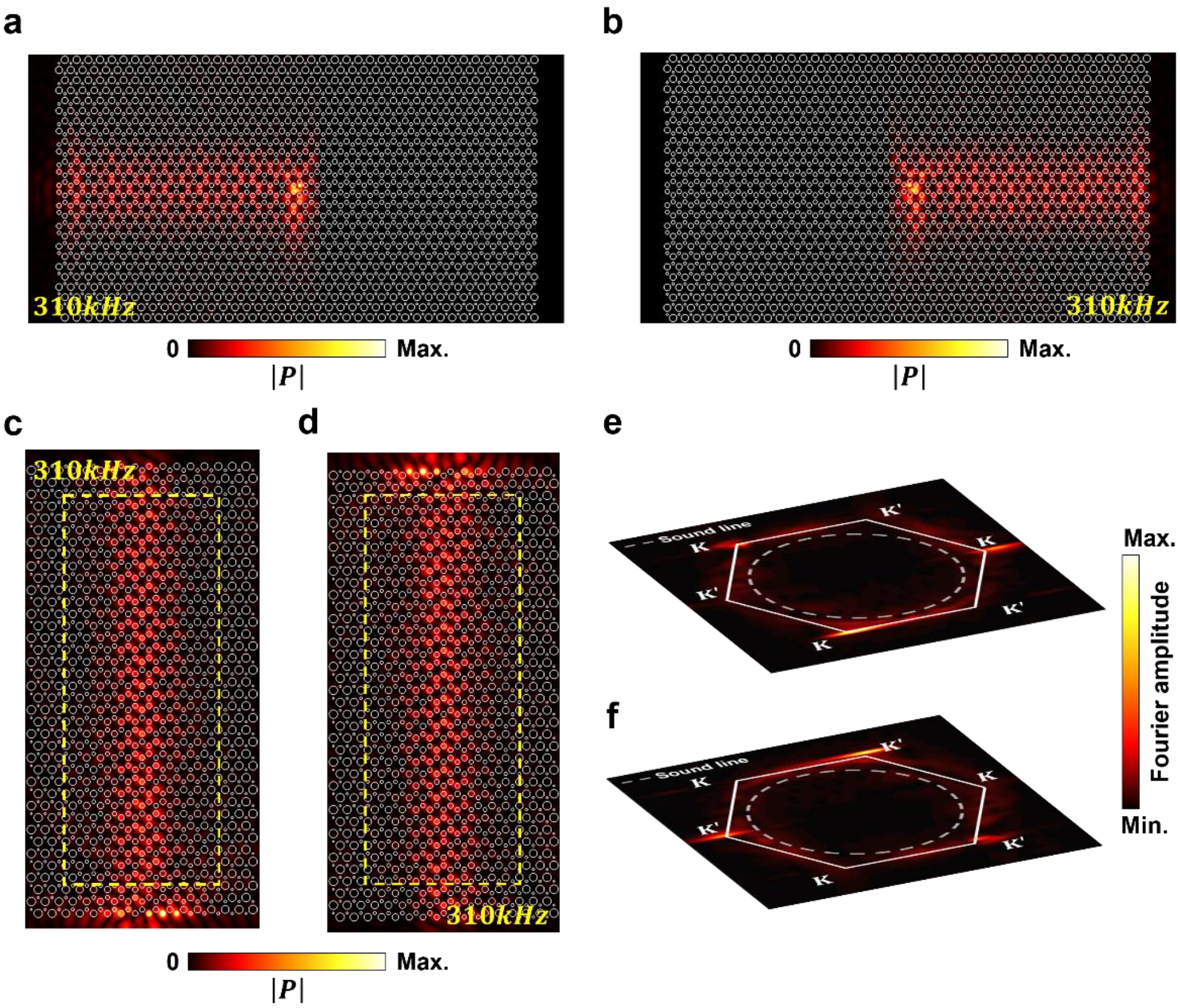

Fig. 5. Visualization of the unidirectional acoustic wave propagation. a) Leftward and (b) rightward one-way propagation of the zeroth LLs excited by a chiral source in the Type I underwater acoustic metamaterial. c) Upward and (d) downward one-way propagation of the zeroth LLs excited by a (c) bottom and (d) top point source in the Type II underwater acoustic metamaterial. e, f) Simulated spatial Fourier spectra of the acoustic pressure field for (c) upward and (d) downward propagation at a frequency of 310 kHz. The data used for the FFT calculation are extracted from the regions outlined by the yellow dashed boxes in (c) and (d). The white hexagons denote the boundaries of the first Brillouin zone, while the gray dashed circles indicate the sound line.

## 4 Experimental validation

Ultrasonic near-field scanning in water is performed to experimentally validate the chiral LLs. The fabricated sample, shown in the lower-left panel of Fig. 6(a), represents the Type I underwater acoustic metamaterial and is formed by periodically arranging the supercell indicated by the dashed box in Fig. 3(a) along the $x$ direction. The sample is manufactured by perforating an aluminum plate (6061-aluminum alloy) using laser cutting. To suppress boundary reflections, the sample edges are coated with a high-stickiness bonding agent (Blu Tack). A piezoelectric actuator (PSt150/2×3/5H, dimensions: 2.1 mm × 3.1 mm × 5 mm) is attached to one facet of the aluminum plate as the experimental excitation source. A needle hydrophone mounted on a 3D motorized stage is used as the detector to perform near-field scanning. A signal generated by a signal generator (Tektronix AFG 3100) and amplified by a power amplifier (Aigtek ATA-2022H) is applied to the actuator, and the excited time-domain ultrasound signals on the other facet of the plate are collected point by point by the hydrophone. Fourier transformation of the time-domain signals yields the field maps in real space at each frequency. Spatial Fourier transforms are then applied to retrieve the band structures in reciprocal space. The signal flow of the experimental setup is given in Fig. 6(a).

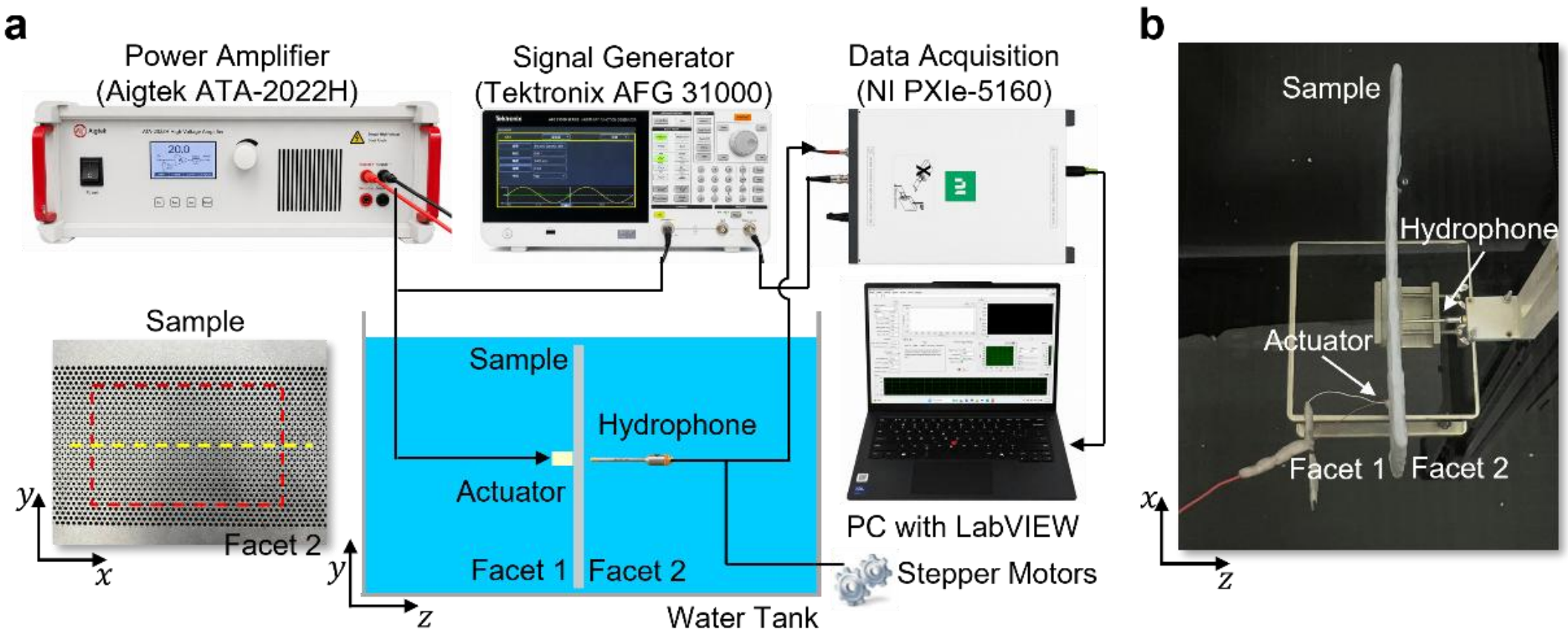

Fig. 6. Experimental setup for ultrasonic near-field scanning in water. a) Schematic of the experimental measurement system. The sample is excited by a piezoelectric actuator attached to facet 1, while the acoustic pressure field on facet 2 is measured point by point using the needle hydrophone. The lower-left panel shows the detailed structure of the fabricated sample, with the red dashed box indicating the scanning region and the yellow dashed line indicating the line scan path. b) Photograph of the fabricated sample and measurement configuration in the water tank.

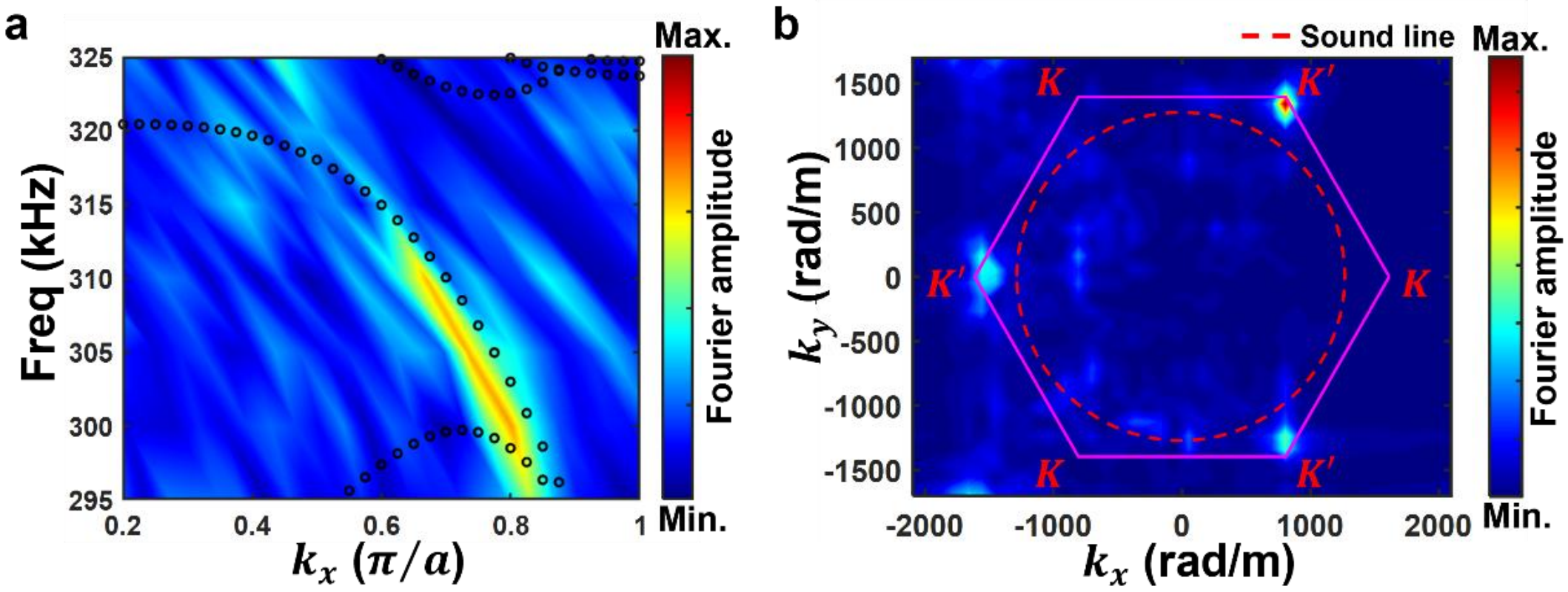

Fig. 7. Experimental validation of the zeroth chiral LL in the Type I underwater acoustic metamaterial. a) Fourier transform of the measured acoustic pressure fields along the yellow dashed line in Fig. 6(a). The color map represents the Fourier amplitude, and the black circles denote the numerically calculated eigenfrequencies. b) Spatial Fourier spectra of the measured acoustic pressure fields within the red dashed box in Fig.6(a) at 310 kHz.

The excitation force is exerted perpendicular to the $x$-$y$ plane to avoid exciting in-plane modes. The actuator is located at the left end of facet 1 of the sample [Fig. 6(b)], and the acoustic pressure is first measured along the yellow dashed line in Fig. 6(a). A Fourier transform of the measured field yields the dispersion of the eigenmodes along the $k_x$ direction, as shown in Fig. 7(a). The experimentally measured zeroth LL, represented by color contours, agrees well with the numerical solutions denoted by black circles. The downward-dispersing zeroth LL exhibits a negative group velocity and corresponds to the $K'$ valley. It is worth noting that the zeroth LL is distinctly different from conventional edge states locked to valleys, featuring the unique characteristics of unidirectional propagating bulk states. Figure 7(b) presents the 2D FFT results of the acoustic pressure field within the red dashed box in Fig. 6(a) at 310 kHz, clearly indicating that the downward-dispersing zeroth LL belongs to the $K'$ valley and further confirming their association.

## 5 Discussion

### *5.1 Robustness*

The zeroth chiral LLs are robust against small perturbations. This robustness is verified by randomly introducing defects into the Type I and Type II underwater acoustic metamaterials, where six additional cylinders with a diameter of 0.4 mm are added, as depicted in Figs. 8(a) and 8(c), with enlarged views shown in the insets.

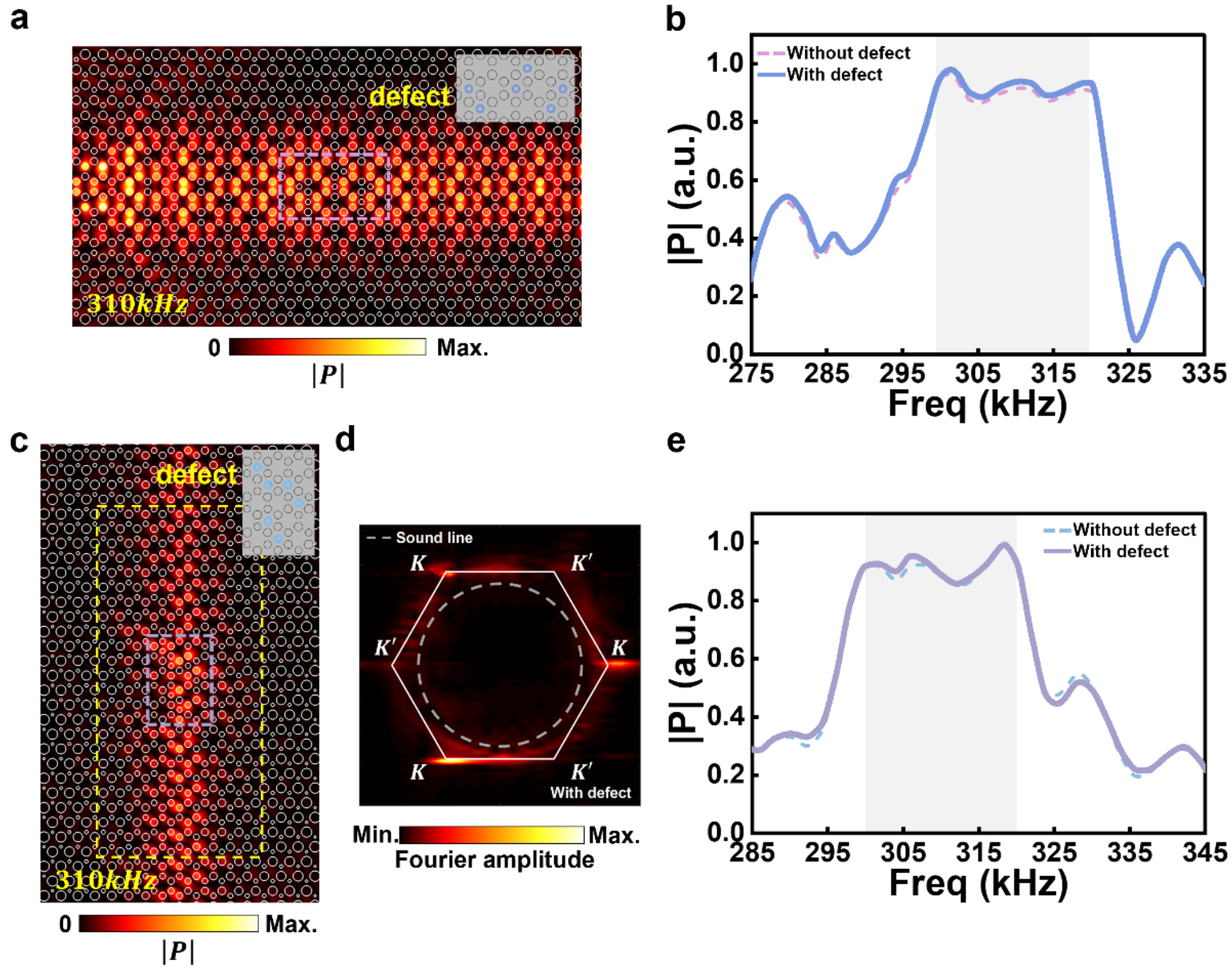

Fig. 8. Robustness validation of chiral LLs. a) Acoustic pressure fields at 310 kHz (left point source) in the Type I underwater acoustic metamaterial with randomly introduced defects. The inset shows an enlarged view of the defect region. b) The transmission spectra of the Type I underwater acoustic metamaterial with and without defects. The gray shaded region indicates the frequency range of the zeroth LL. c) Acoustic pressure fields at 310 kHz (bottom point source) in the Type II underwater acoustic metamaterial with randomly introduced defects. The inset shows an enlarged view of the defect region. d) Spatial Fourier spectra of the acoustic pressure fields within the yellow dashed box in (c). e) The transmission spectra of the Type II underwater acoustic metamaterial with and without defects. The gray shaded region indicates the frequency range of the zeroth LL.

A point source is placed at the middle of the left boundary of the Type I metamaterial and at the middle of the bottom boundary of the Type II metamaterial. The localization characteristics of the zeroth LL and the large-area transport of acoustic energy remain nearly unaffected by the introduced defects. Figures 8(b) and 8(e) present the transmission spectra of the Type I and Type II underwater acoustic metamaterials with and without defects. Enhanced transmission is observed within the shaded frequency ranges of the zeroth LLs, confirming their relatively broad bandwidths. The close agreement between the spectra before and after introducing defects demonstrates the robustness of the chiral zeroth LLs. The 2D FFT results of the acoustic pressure fields in the Type II metamaterials with defects are shown in Fig. 8(d). The results confirm that, even in the presence of defects, the wave vectors of the propagating acoustic waves remain locked around the $K$ or $K'$ valleys, validating the excellent robustness of the zeroth LLs against defects.

*5.2 Flexible underwater ultrasound manipulations*

The topological origin of chiral LLs enables a series of exotic wave manipulations that are difficult to achieve in conventional metamaterials, including beam splitting in the Type I underwater acoustic metamaterial. The overall structure of the designed beam splitter is illustrated in Fig. 9(a), with the propagation paths of underwater ultrasound marked by blue shaded regions. The unit cells with zero Dirac mass are initially arranged along the horizontal midline of the phononic crystal plate. These unit cells then gradually separate upward and downward and finally terminate at two well-separated positions (see Supplementary Note 3 for the detailed geometrical parameters). When an excitation is applied at the center of the left boundary of the phononic crystal plate, the acoustic waves gradually split during propagation, as depicted in Fig. 9(b). This mechanism of constructing underwater beam splitters highlights the promising potential of underwater acoustic metamaterials in spatial diversity and spatial multiplexing.

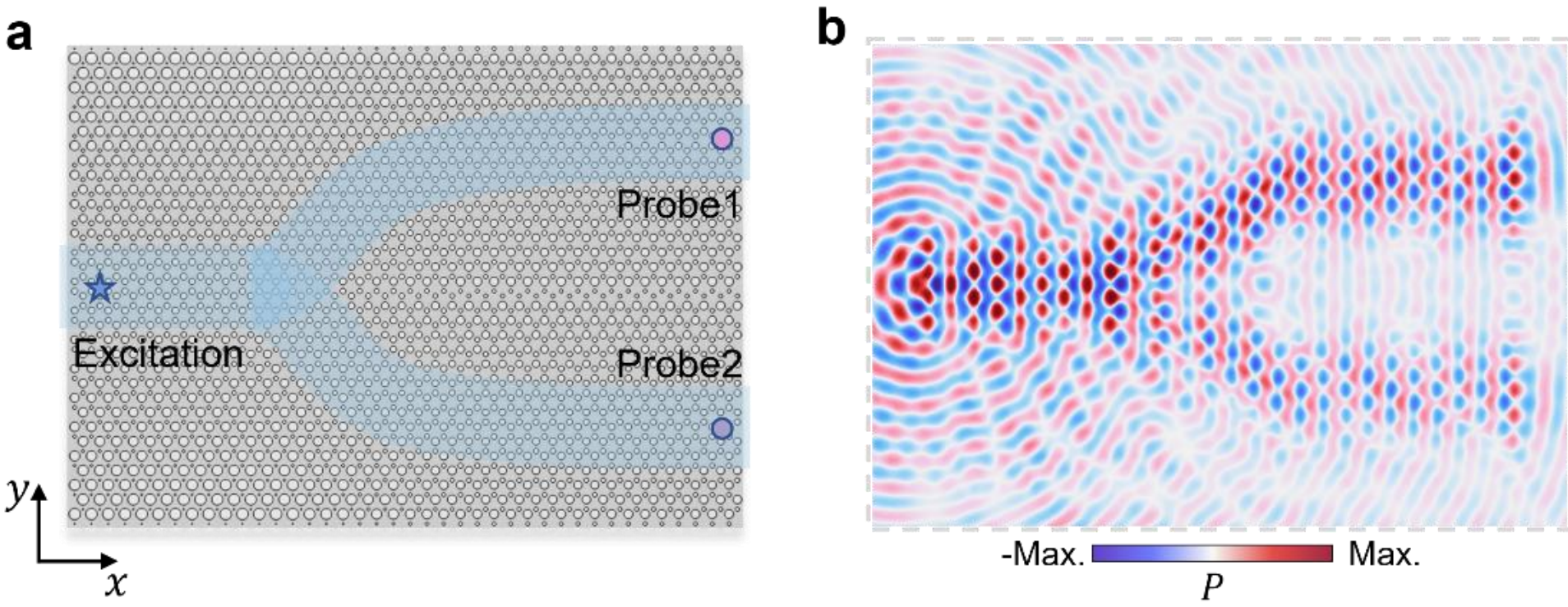

Fig. 9. Demonstration of underwater acoustic energy splitting based on the Type I underwater acoustic metamaterial. a) Designed phononic crystal plate for acoustic energy splitting. The excitation source is placed at the left side of the sample, and the blue shaded regions indicate the splitting paths of underwater ultrasound. b) The acoustic pressure field in the Type I underwater acoustic metamaterial (left point source).

In addition, the localized nature of the zeroth LLs around the central unit cells of the system and their robustness against defects demonstrated above provide a unique opportunity, which enables acoustic energy transport along prescribed propagation paths by strategically shifting the supercells leftward or rightward accordingly. As illustrated in Fig. 10(a), the Type II underwater acoustic metamaterial is composed of five blocks, each containing four consecutive supercells. Each block is shifted to the left or right by a certain distance, forming a curved propagation path for acoustic waves. The acoustic pressure field in Fig. 10(b) shows that the acoustic waves propagate smoothly along the designed path, with the energy confined near the center of the path. A path with a larger steering angle is further constructed by increasing the lateral shifts of the supercell blocks, as shown in Fig. 10(c). The acoustic field remains localized along the customized path, indicating that underwater acoustic waves can be routed through prescribed trajectories. This feature provides a promising approach for transmitting underwater acoustic waves along desired paths.

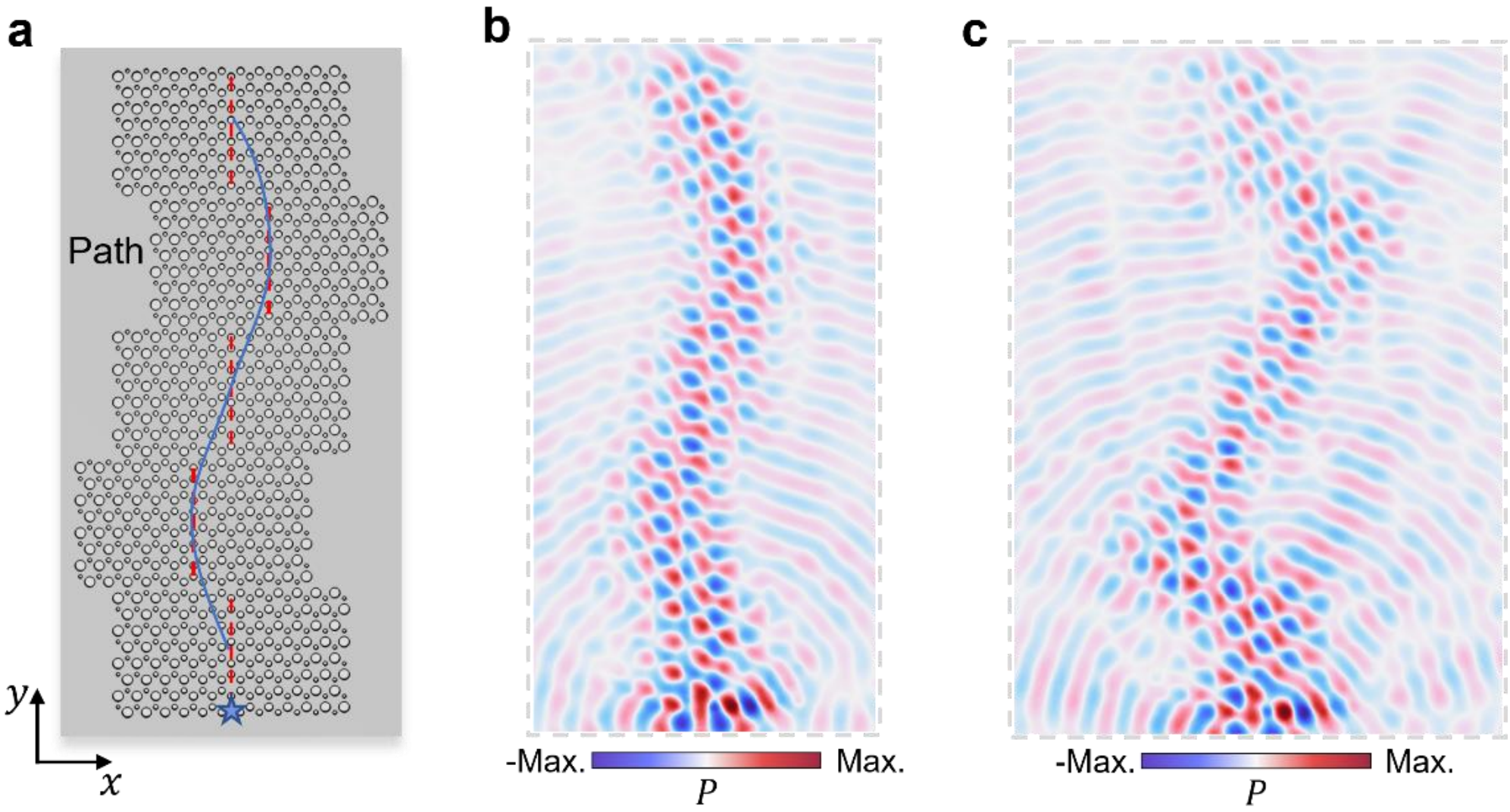

Fig. 10. Demonstration of underwater acoustic energy transport along prescribed paths. a) Designed phononic crystal plate with a targeted propagation path formed by laterally shifting the supercells. The blue curve indicates the designed transport path, and the red dashed lines mark the shifted supercell blocks. b) The acoustic pressure field in the Type II underwater acoustic metamaterial (bottom point source). c) The acoustic pressure field for a path with a larger steering angle (larger lateral shifts).

### *5.3 Dual-band chiral Landau levels*

A novel phenomenon emerges when the dimensions of the underwater acoustic metamaterial are scaled down to a certain extent: dual-band chiral LLs can be realized. Starting from the unit cell shown in Fig. 2(a), the geometric dimensions are approximately scaled down, while the plate thickness is increased. For a lattice constant of $a = 1.212$ mm, circular hole diameters of $d_A = d_B = 0.5$ mm, and a plate thickness of $l = 2$ mm, two Dirac cones appear at the $K/K'$ points of the Brillouin zone around 317 kHz and 619 kHz, as indicated by the dashed curves in Figs. 11(a) and (b). By introducing a circular hole diameter difference ($\Delta d$), the inversion symmetry of the unit cell is broken. For $d_A = 0.45$ mm and $d_B = 0.55$ mm, bandgaps are opened at both Dirac points, as shown by the solid curves in Figs. 11(a) and (b), providing the basis for realizing dual-band chiral LLs.

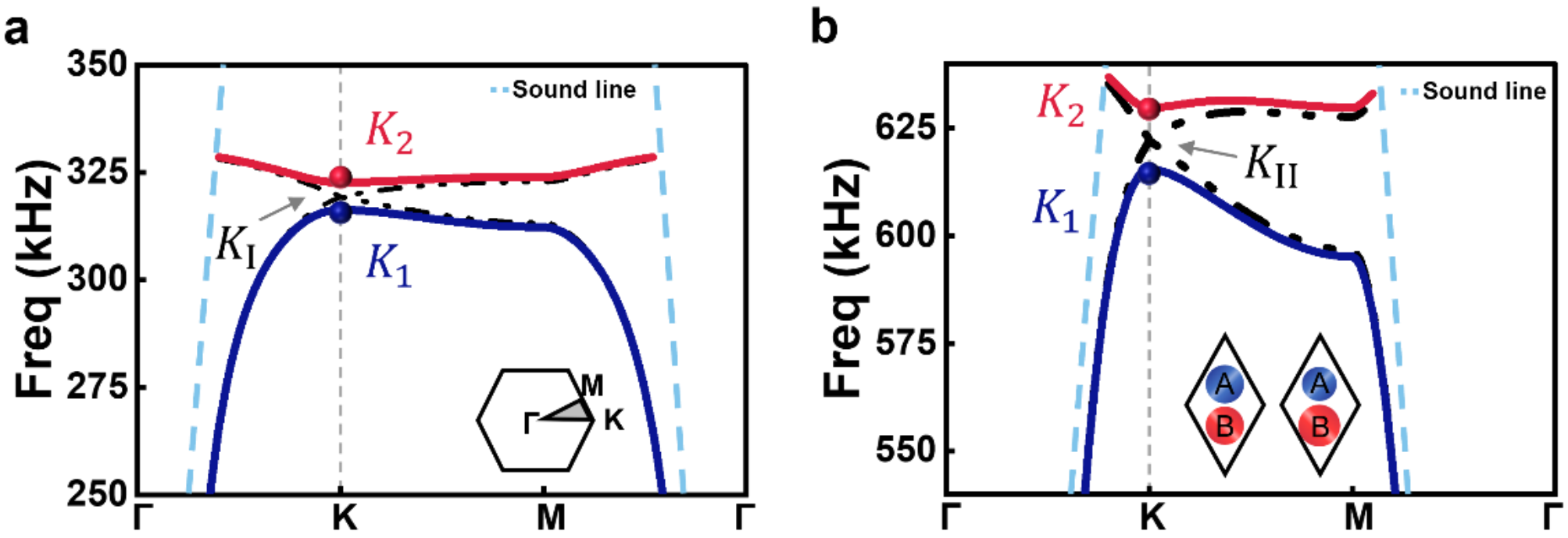

Fig. 11. Band structures of the small-scale unit cell. a, b) The band structures around the first ($K_{\mathrm{I}}$) and second ($K_{\mathrm{II}}$) Dirac cones, respectively, for the unit cells with $d_A = d_B = 0.5$ mm (Dirac cones, dashed lines) and $d_A = 0.45$ mm, $d_B = 0.55$ mm (opened bandgaps, solid lines).

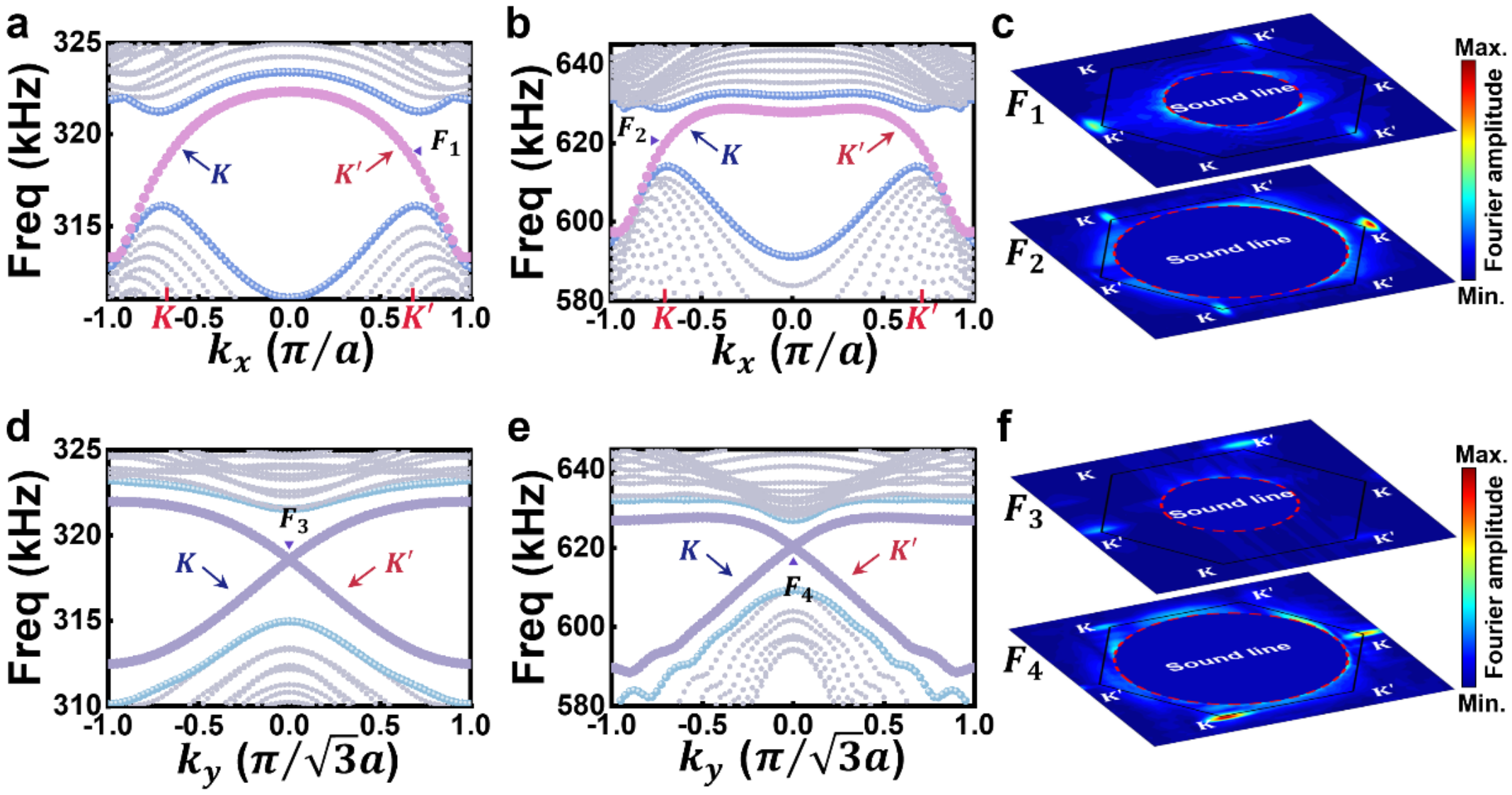

Fig. 12. Dual-band chiral LLs in a small-scale underwater acoustic metamaterial. a, b) The projected band structures of the supercell near 320 kHz and 620 kHz in the Type I underwater acoustic metamaterial. c) Spatial Fourier spectra of the acoustic pressure fields corresponding to $F_1$ and $F_2$. d, e) The projected band structures of the supercell near 320 kHz and 620 kHz in the Type II underwater acoustic metamaterial. f) Spatial Fourier spectra of the acoustic pressure fields corresponding to $F_3$ and $F_4$.

Guided by the theoretical derivations in Section 2, underwater acoustic metamaterials based on the scaled unit cell are designed to generate in-plane PMFs along the $x$ and $y$ directions, enabling dual-band chiral LLs. Figures 12(a) and 12(b) show the projected band structures of the supercell for the dual-band Type I underwater acoustic metamaterial induced by $B_x$. Chiral LLs emerge in both frequency bands, each with a certain bandwidth. At the frequencies marked by $F_1$ and $F_2$, the acoustic pressure fields are excited to propagate rightward and leftward, respectively, and their 2D Fourier spectra are shown in Fig. 12(c). The Fourier amplitudes are concentrated around the $K$ or $K'$ valleys, confirming the valley-locked nature of the propagating bulk states. Similar dual-band behavior is observed in the Type II underwater acoustic metamaterial induced by $B_y$, as shown in Figs. 12(d) and 12(e). The states $F_3$ and $F_4$ correspond to downward and upward propagation, respectively, while their Fourier spectra in Fig. 12(f) remain concentrated around the $K$ or $K'$ valleys. These results verify that dual-band chiral LLs support unidirectional bulk propagation in both Type I and Type II underwater acoustic metamaterials.

## 6 Conclusion

In conclusion, broadband chiral LLs of underwater SSAWs, induced by $x$-directed and $y$-directed in-plane PMFs ($B_x$ and $B_y$), have been realized in open fluid-solid coupled acoustic metamaterials. These two PMFs are constructed by linearly opening local bandgaps at the Dirac points and introducing distinct position-dependent effective mass terms. The resulting chiral LLs support valley-dependent unidirectional transmission of underwater SSAWs and exhibit robustness against defects. The experimentally measured dispersion spectra in the Type I metamaterial agree well with numerical calculations. Notably, these chiral LLs are implemented along an exposed fluid-solid interface without relying on confinement within an enclosed acoustic waveguide. Beyond confirming the existence of underwater chiral LLs, the proposed framework enables flexible control of underwater ultrasound through the modulation of in-plane PMFs, including beam splitting and wave transmission along customized paths. The realization of dual-band chiral LLs in a small-scale underwater acoustic metamaterial further demonstrates the scalability of this mechanism and its potential applications to multiband underwater acoustic information processing, underwater detection, underwater communication, and underwater acoustic energy harvesting.

## CRediT authorship contribution statement

**Jiao Shen:** Conceptualization, Methodology, Software, Investigation, Writing – original draft. **Zhiyong Chang:** Methodology, Validation, Supervision. **Xinzong Wang:** Software, Investigation, Validation. **Cheng Lin:** Investigation, Validation, Visualization. **Tuo Liu:** Methodology, Validation, Writing – review & editing. **Xiaoxiao Wu:** Formal analysis, Methodology, Writing – review & editing. **Yifan Zhu:** Validation, Resources, Writing – review & editing. **Haiyan Fan:** Supervision, Project administration, Funding acquisition, Writing – review & editing. **Hui Zhang:** Conceptualization, Supervision, Project administration, Funding acquisition, Writing – review & editing.

## Declaration of competing interest

The authors declare that they have no known competing financial interests or personal relationships that could have appeared to influence the work reported in this paper.

## Acknowledgments

This work was supported by the National Natural Science Foundation of China (Grant Nos. 52501418 and 12574493), the Natural Science Foundation of Jiangsu Province (Grant No. BK20251266), and the Advanced Ocean Institute of Southeast University, Nantong (MP202601).

## References

[1] E. Dong, P. Cao, J. Zhang, S. Zhang, N.X. Fang, Y. Zhang, Underwater acoustic metamaterials, Natl. Sci. Rev. 10 (6) (2023) nwac246.
[2] H.-T. Zhou, W.-X. Fu, X.-S. Li, Y.-F. Wang, Y.-S. Wang, Loosely coupled reflective impedance metasurfaces:

Precise manipulation of waterborne sound by topology optimization, Mech. Syst. Signal Process. 177 (2022) 109228.
[3] Y. Chen, M. Zheng, X. Liu, Y. Bi, Z. Sun, P. Xiang, J. Yang, G. Hu, Broadband solid cloak for underwater acoustics, Phys. Rev. B. 95 (18) (2017) 180104.
[4] Y. Bi, H. Jia, Z. Sun, Y. Yang, H. Zhao, J. Yang, Experimental demonstration of three-dimensional broadband underwater acoustic carpet cloak, Appl. Phys. Lett. 112 (22) (2018) 223502.
[5] Z. Sun, X. Sun, H. Jia, Y. Bi, J. Yang, Quasi-isotropic underwater acoustic carpet cloak based on latticed pentamode metafluid, Appl. Phys. Lett. 114 (9) (2019) 094101.
[6] T. Liang, M. He, H.-W. Dong, L. Xia, X. Huang, Ultrathin waterborne acoustic metasurface for uniform diffuse reflections, Mech. Syst. Signal Process. 192 (2023) 110226.
[7] E. Dong, Y. Zhang, Z. Song, T. Zhang, C. Cai, N.X. Fang, Physical modeling and validation of porpoises' directional emission via hybrid metamaterials, Natl. Sci. Rev. 6 (5) (2019) 921–928.
[8] E. Dong, Z. Song, Y. Zhang, S. Ghaffari Mosanenzadeh, Q. He, X. Zhao, N.X. Fang, Bioinspired metagel with broadband tunable impedance matching, Sci. Adv. 6 (44) (2020) eabb3641.
[9] Z. Song, W. Ou, E. Dong, J. Zhang, Q. Xie, C. Zhang, M. Bai, T.A. Mooney, Y. Zhang, Physical implementation of dolphin biosonar to facilitate ultrasound control, Appl. Phys. Lett. 117 (17) (2020) 173701.
[10] A. Marzo, S.A. Seah, B.W. Drinkwater, D.R. Sahoo, B. Long, S. Subramanian, Holographic acoustic elements for manipulation of levitated objects, Nat. Commun. 6 (1) (2015) 8661.
[11] X. Zhu, K. Li, P. Zhang, J. Zhu, J. Zhang, C. Tian, S. Liu, Implementation of dispersion-free slow acoustic wave propagation and phase engineering with helical-structured metamaterials, Nat. Commun. 7 (1) (2016) 11731.
[12] S. Jiménez-Gambín, N. Jiménez, J.M. Benlloch, F. Camarena, Generating Bessel beams with broad depth-of-field by using phase-only acoustic holograms, Sci. Rep. 9 (1) (2019) 20104.
[13] Y.-X. Shen, Y.-G. Peng, F. Cai, K. Huang, D.-G. Zhao, C.-W. Qiu, H. Zheng, X.-F. Zhu, Ultrasonic super-oscillation wave-packets with an acoustic meta-lens, Nat. Commun. 10 (1) (2019) 3411.
[14] H. Zhang, W. Zhang, Y. Liao, X. Zhou, J. Li, G. Hu, X. Zhang, Creation of acoustic vortex knots, Nat. Commun. 11 (1) (2020) 3956.
[15] J.-W. Kim, G. Hwang, S.-J. Lee, S.-H. Kim, S. Wang, Three-dimensional acoustic metamaterial Luneburg lenses for broadband and wide-angle underwater ultrasound imaging, Mech. Syst. Signal Process. 179 (2022) 109374.
[16] C. He, Z. Li, X. Ni, X.-C. Sun, S.-Y. Yu, M.-H. Lu, X.-P. Liu, Y.-F. Chen, Topological phononic states of underwater sound based on coupled ring resonators, Appl. Phys. Lett. 108 (3) (2016) 031904.
[17] J. Lu, H. Zhai, Q. Shi, H. Zhang, D. Zhao, Y. Yao, Y. Huang, L. Luo, X. Zhang, Underwater broadband topological slow-wave in a spin-Chern phononic crystal, Phys. Rev. B. 112 (1) (2025) 014104.
[18] Y. Shen, C. Qiu, X. Cai, L. Ye, J. Lu, M. Ke, Z. Liu, Valley-projected edge modes observed in underwater sonic crystals, Appl. Phys. Lett. 114 (2) (2019) 023501.
[19] S. Zheng, G. Duan, B. Xia, Underwater acoustic positioning based on valley-chirality locked beam of sonic system, Int. J. Mech. Sci. 174 (2020) 105463.
[20] J. Liang, R. Zheng, Z. Lu, J. Pan, J. Lu, W. Deng, M. Ke, X. Huang, Z. Liu, Corner states and particle trapping in waterborne acoustic crystals, Appl. Phys. Lett. 124 (10) (2024) 101704.
[21] J. Liang, H. Liu, R. Zheng, J. Wu, J. Lu, W. Deng, M. Ke, X. Huang, Z. Liu, Multiple corner states and particle trapping in higher-order topological waterborne acoustic crystals, Phys. Rev. Appl. 23 (2) (2025) 024024.
[22] Z. Chen, G. Wang, Y. Mao, C.W. Lim, New metamaterial mathematical modeling of acoustic topological insulators via tunable underwater local resonance, Appl. Math. Model. 108 (2022) 258–274.
[23] X. Wu, H. Fan, T. Liu, Z. Gu, R.-Y. Zhang, J. Zhu, X. Zhang, Topological phononics arising from fluid-solid interactions, Nat. Commun. 13 (1) (2022) 6120.
[24] L. Luo, L. Ye, H. He, J. Lu, M. Ke, Z. Liu, Experimental observation of the topological interface state in a one-dimensional waterborne acoustic waveguide, Phys. Rev. Appl. 20 (4) (2023) 044008.
[25] H.-W. Dong, S.-D. Zhao, P. Xiang, B. Wang, C. Zhang, L. Cheng, Y.-S. Wang, D. Fang, Porous-Solid Metaconverters for Broadband Underwater Sound Absorption and Insulation, Phys. Rev. Appl. 19 (4) (2023) 044074.
[26] H. Zou, L. Su, Y. Zhang, M. Zhang, W. Yu, X. Wang, X. Xia, H. Chen, X. Zhang, A. Zhao, A novel broadband underwater sound absorption metastructure with multi-oscillators, Int. J. Mech. Sci. 271 (2024) 109137.
[27] X. Fang, N. Wang, B. Yan, D. Li, Y. Li, W. Wang, W. Wu, Nonlocal metagrating for low-frequency and broadband underwater sound absorption, Int. J. Mech. Sci. 286 (2025) 109917.
[28] Z. Zheng, H. Yang, M. Wang, Y. Wang, J. Zhong, W. Ma, C. Liang, J. Wen, X. Chen, Realization of lightweight

and pressure-resistant sandwich metasurfaces for underwater sound absorption through topology optimization, Mech. Syst. Signal Process. 224 (2025) 112205.
[29] K. v. Klitzing, G. Dorda, M. Pepper, New Method for High-Accuracy Determination of the Fine-Structure Constant Based on Quantized Hall Resistance, Phys. Rev. Lett. 45 (6) (1980) 494–497.
[30] R.B. Laughlin, Quantized Hall conductivity in two dimensions, Phys. Rev. B. 23 (10) (1981) 5632–5633.
[31] D.J. Thouless, M. Kohmoto, M.P. Nightingale, M. den Nijs, Quantized Hall Conductance in a Two-Dimensional Periodic Potential, Phys. Rev. Lett. 49 (6) (1982) 405–408.
[32] Q. Niu, D.J. Thouless, Y.-S. Wu, Quantized Hall conductance as a topological invariant, Phys. Rev. B. 31 (6) (1985) 3372–3377.
[33] N. Levy, S.A. Burke, K.L. Meaker, M. Panlasigui, A. Zettl, F. Guinea, A.H.C. Neto, M.F. Crommie, Strain-Induced Pseudo–Magnetic Fields Greater Than 300 Tesla in Graphene Nanobubbles, Science. 329 (5991) (2010) 544–547.
[34] F. Guinea, M.I. Katsnelson, A.K. Geim, Energy gaps and a zero-field quantum Hall effect in graphene by strain engineering, Nat. Phys. 6 (1) (2010) 30–33.
[35] M.C. Rechtsman, J.M. Zeuner, A. Tünnermann, S. Nolte, M. Segev, A. Szameit, Strain-induced pseudomagnetic field and photonic Landau levels in dielectric structures, Nat. Photonics. 7 (2) (2013) 153–158.
[36] D.-B. Zhang, G. Seifert, K. Chang, Strain-Induced Pseudomagnetic Fields in Twisted Graphene Nanoribbons, Phys. Rev. Lett. 112 (9) (2014) 096805.
[37] H. Abbaszadeh, A. Souslov, J. Paulose, H. Schomerus, V. Vitelli, Sonic Landau Levels and Synthetic Gauge Fields in Mechanical Metamaterials, Phys. Rev. Lett. 119 (19) (2017) 195502.
[38] Z. Yang, F. Gao, Y. Yang, B. Zhang, Strain-Induced Gauge Field and Landau Levels in Acoustic Structures, Phys. Rev. Lett. 118 (19) (2017) 194301.
[39] M. Bellec, C. Poli, U. Kuhl, F. Mortessagne, H. Schomerus, Observation of supersymmetric pseudo-Landau levels in strained microwave graphene, Light Sci. Appl. 9 (1) (2020) 146.
[40] X. Wen, C. Qiu, Y. Qi, L. Ye, M. Ke, F. Zhang, Z. Liu, Acoustic Landau quantization and quantum-Hall-like edge states, Nat. Phys. 15 (4) (2019) 352–356.
[41] O. Jamadi, E. Rozas, G. Salerno, M. Milićević, T. Ozawa, I. Sagnes, A. Lemaître, L. Le Gratiet, A. Harouri, I. Carusotto, J. Bloch, A. Amo, Direct observation of photonic Landau levels and helical edge states in strained honeycomb lattices, Light Sci. Appl. 9 (1) (2020) 144.
[42] W. Wang, W. Gao, X. Chen, F. Shi, G. Li, J. Dong, Y. Xiang, S. Zhang, Moir\'e Fringe Induced Gauge Field in Photonics, Phys. Rev. Lett. 125 (20) (2020) 203901.
[43] H. Xue, Q. Wang, B. Zhang, Y.D. Chong, Non-Hermitian Dirac Cones, Phys. Rev. Lett. 124 (23) (2020) 236403.
[44] M. Yan, Pseudomagnetic Fields Enabled Manipulation of On-Chip Elastic Waves, Phys. Rev. Lett. 127 (13) (2021).
[45] Z.-M. Yu, Y. Yao, S.A. Yang, Predicted Unusual Magnetoresponse in Type-II Weyl Semimetals, Phys. Rev. Lett. 117 (7) (2016) 077202.
[46] D.I. Pikulin, A. Chen, M. Franz, Chiral Anomaly from Strain-Induced Gauge Fields in Dirac and Weyl Semimetals, Phys. Rev. X. 6 (4) (2016) 041021.
[47] V. Peri, M. Serra-Garcia, R. Ilan, S.D. Huber, Axial-field-induced chiral channels in an acoustic Weyl system, Nat. Phys. 15 (4) (2019) 357–361.
[48] H. Jia, R. Zhang, W. Gao, Q. Guo, B. Yang, J. Hu, Y. Bi, Y. Xiang, C. Liu, S. Zhang, Observation of chiral zero mode in inhomogeneous three-dimensional Weyl metamaterials, Science. 363 (6423) (2019) 148–151.
[49] K. Li, X. Zhang, Z. Zhang, Y. Cheng, X. Liu, Observation of chiral Landau levels in a synthetic acoustic Weyl semimetal, Commun. Phys. 8 (1) (2025) 133.
[50] H. Jia, M. Wang, S. Ma, R.-Y. Zhang, J. Hu, D. Wang, C.T. Chan, Experimental realization of chiral Landau levels in two-dimensional Dirac cone systems with inhomogeneous effective mass, Light Sci. Appl. 12 (1) (2023) 165.
[51] R. Barczyk, L. Kuipers, E. Verhagen, Observation of Landau levels and chiral edge states in photonic crystals through pseudomagnetic fields induced by synthetic strain, Nat. Photonics. 18 (6) (2024) 574–579.
[52] Y. Yang, J. Zhang, B. Yang, S. Liu, W. Zhang, X. Shen, L. Shi, Z.H. Hang, Photonic Dirac waveguide in inhomogeneous spoof surface plasmonic metasurfaces, Nanophotonics. 13 (20) (2024) 3847–3854.
[53] Y. Liu, K. Li, W. Liu, Z. Zhang, Y. Cheng, X. Liu, Observation of chiral Landau levels in two-dimensional acoustic system, Quantum Front. 3 (1) (2024) 26.
[54] Z. Cui, C. Wu, Q. Wei, M. Yan, G. Chen, On-Chip Elastic Wave Manipulations Based on Synthetic Dimension,

Phys. Rev. Lett. 133 (25) (2024) 256602.

[55] S. Li, P.G. Kevrekidis, J. Yang, Emergence of elastic chiral Landau levels and snake states, Phys. Rev. B. 109 (18) (2024) 184109.

[56] Y. Chen, Z. Lan, L. Fan, S. Liang, L. An, J. Zhu, Z. Su, Realization of broadband ultrasonic chiral Landau levels in an elastic metamaterial, Phys. Rev. B. 111 (18) (2025) 184303.

[57] Y. Chen, Z. Lan, S. Liang, S.C. Cheung, L. Fan, J. Zhu, Z. Su, Visualization of flexible large-area acoustic energy conveying enabled by valley-dependent Landau bulk modes, Sci. China Phys. Mech. Astron. 68 (9) (2025) 294311.

[58] L. Fan, Z. Lan, Y. Chen, J. Zhu, Z. Su, Dual-band flexible large-area ultrasonic energy conveying via elastic chiral Landau levels, Mater. Horiz. 12 (16) (2025) 6334–6341.

[59] S. Huang, L. Chen, Z. Lan, Q.C. Yin, S.L. Qin, J.W. You, T.J. Cui, Synthetic In-Plane Magnetic Fields in Topological Plasmonic Insulator with Detachable Identical Meta-Atoms, Laser Photonics Rev. 19 (23) (2025) e00280.

[60] Y. Yuan, Y. Xu, L. Zhao, Q. He, S. Sun, S. Ma, L. Zhou, Supersymmetric Landau Levels in Subwavelength Type-I Dirac Metasurfaces, Phys. Rev. Lett. 136 (2) (2026) 023802.

[61] C. Lu, O. You, Y. Zheng, S. Ma, Q. Yang, H.-C. Chan, Z. Wang, W. Ma, Y.-C. Liu, S. Zhang, Realizing Landau levels with open equifrequency contours around band singularities, Nat. Commun. 17 (1) (2026) 4889.